\documentclass[preprint,showpacs,preprintnumbers,amsmath,amssymb]{revtex4}

\usepackage{graphicx}
\usepackage{dcolumn}
\usepackage{bm}

\begin{document}

\preprint{}

\title{Established Clustering Procedures for Network Analysis}

\author{Paul B. Slater}%
\email{slater@kitp.ucsb.edu}
\affiliation{%
ISBER, University of California, Santa Barbara, CA 93106\\
}%
\date{\today}

\begin{abstract}
In light of the burgeoning interest in
network analysis in the new millenium, we
bring to the attention of {\it contemporary} 
network theorists, a
two-stage {\it double}-standarization and hierarchical clustering 
(single-linkage-like) procedure devised in 1974. In its many applications 
over the next decade--primarily to the migration flows between 
geographic subdivisions within
nations--the presence was often revealed of ``hubs''. 
These are, 
typically, ``cosmopolitan/non-provincial'' areas--such as the French capital, 
Paris--which send and receive people {\it relatively} broadly 
across their respective  nations. Additionally, this two-stage 
procedure--which ``might very well be the most successful application
of cluster analysis'' (R. C. Dubes)--has detected many
(physically or socially) isolated groups 
(regions)
of areas, such as those forming the southern
islands, Shikoku and Kyushu, of Japan, the Italian islands 
of Sardinia and Sicily, 
and the New England region of the 
United States. Further, we discuss a (complementary)
approach developed in 1976, 
involving the application of the {\it max-flow/min-cut} theorem
to {\it raw/non-standardized} flows.
\end{abstract}

\pacs{Valid PACS 02.10.Ox, 89.65.-s}
\keywords{networks, hubs, clusters, internal migration, flows, strong
components, graph theory, 
hierarchical cluster analysis, dendrograms, cosmopolitan areas}

\maketitle
A. L. Barab{\' a}si, in his 
recent popular book, ``Linked'', asserts that the emergence 
of {\it hubs} in networks is a surprising phenomenon that is ``forbidden by both the Erd{\"o}s-R{\'e}nyi and Watts-Strogatz models" \cite[p. 63]{linked} 
\cite[Chap. 8]{siegfried}.
Here, we indicate an analytical framework introduced
in 1974 that the distinguished computer scientist 
R. C. Dubes, in a review of \cite{tree}, has asserted 
``might very well be the most successful application 
of cluster analysis'' \cite[p. 142]{dubes}. It has proved insightful
in revealing--among other interesting relationships--hub-like structures 
in networks of (weighted, directed) internodal flows. 
In the recent resurgence of interest in network analysis, this methodology
may have been overlooked,
as many of its uses had been
reported in the 1970's and 1980's, in journals 
outside of the strictly mathematical and physical literature 
\cite{japan,france,winchester,science,partial,IO,college,gentileschi,SEAS,qq} 
(as well as in the research institute monographs \cite{tree,county,tree2}, 
widely distributed to academic libraries).

Though the principal procedure under discussion here is applicable 
in a wide variety of social-science 
settings \cite{tree,dubes}, it has been largely 
used, in a demographic context, to study the 
{\it internal migration} tables published at regular periodic intervals by most of the nations of the world. 
These tables can be thought of as $N \times N$ (square) matrices, the entries ($m_{ij}$) of which are the number of people who
lived in geographic subdivision $i$ at time $t$ and $j$ at 
time $t+1$. (Some tables--but not all--have diagonal entries, $m_{ii}$, which may represent the number of people who did move within
area $i$, or simply those who lived in $i$ at $t$ and $t+1$.)

In the {\it first} step of the 
analytical procedure employed, the rows and columns of the 
table of flows 
are alternately  (biproportionally \cite{bacharach}) scaled 
to sum to a fixed number (say 1). Under broad conditions--to be discussed 
below--convergence occurs to a ``doubly-stochastic" 
(bistochastic) 
table, with row and column sums all 
simultaneously equal to 1 \cite{mosteller,louck,CSBZ,unistochastic}. 
The purpose of the scaling is to 
remove overall (marginal) effects of size, and focus on relative, 
interaction  effects. 
The {\it cross-product ratios} 
({\it relative odds}), $\frac{m_{ij} m_{kl}}{m_{il} m_{kj}}$, 
measures of association, are left {\it invariant}. 
Additionally, the entries of the
doubly-stochastic table provide 
{\it maximum  entropy} estimates of the original
flows, given the row and column constraints \cite{eriksson}.

For large {\it sparse} 
flow tables, only the nonzero entries, together with their
row and column coordinates are needed. Row and column (biproportional) 
multipliers can be iteratively computed by sequentially accessing the nonzero
cells \cite{parlett}. If the table is ``critically sparse'', various convergence difficulties may occur. Nonzero entries that are ``unsupported''--that is, not part of a set of $N$ nonzero entries, no two in the same row and 
column-- may converge to zero and/or the 
biproportional multipliers may not converge \cite[p. 19]{tree} \cite{sinkhorn} \cite[p. 171]{mirsky}.
(The scaling was successfully implemented with a
$3,140 \times 3,140$ 1965-70 intercounty migration table for
the United States \cite{county,partial}--as well as for a more aggregate
$510 \times 510$ table for the US \cite{SEAS}. {\it Smoothing} procedures
can be used to modify the zero-nonzero structure 
of a flow table, particularly 
if it is critically sparse \cite{simonoff,boundary}.)
The ``first strongly polynomial-time algorithm for matrix scaling'' 
was reported in \cite{linial}.
 
In the {\it second} step of the procedure, the doubly-stochastic matrix 
is converted to a series of {\it directed} 
(0,1) graphs (digraphs), by applying thresholds to its entries. 
As the thresholds   are 
progressively lowered, larger and larger {\it strong components} 
(a directed path existing from any member of a component 
to any other) of the resulting 
graphs are found. This process 
(a simple variant of well-known single-linkage [nearest neighbor]
clustering) can be represented by the familiar dendrogram 
or tree diagram used in 
hierarchical cluster analysis and cladistics/phylogeny (cf. \cite{ozawa}).

A FORTRAN implementation of the two-stage process was given in 
\cite{leusmann}, as well as one in the SAS (Statistical Analysis System) 
framework \cite{chilko}. 
The noted computer scientist R. E. Tarjan 
\cite{schwartz} devised an $O(M (\log{N})^2)$
algorithm \cite{tarjan} and, then, a further improved $O(M (\log{N}))$ 
method \cite{tarjan2}, 
where $N$ is the number of nodes and $M$ the number of edges of 
a directed graph. (These substantially improved upon 
the earlier works \cite{leusmann,chilko}, 
which 
required the 
computations of {\it transitive closures} of graphs, and were 
$O(M N)$ in nature.) A FORTRAN coding--involving 
linked lists--of the improved Tarjan 
algorithm \cite{tarjan2} was presented
in \cite{tarjanslater}, and applied in the US intercounty study \cite{county}. 
If the graph-theoretic (0,1) structure of the network under study 
is {\it not} strongly connected 
\cite{hartfiel}, {\it independent} analyses of 
the subsystems of the network is appropriate.

The {\it goodness-of-fit} of the dendrogram generated 
to the doubly-stochastic table itself
can be evaluated--and possibly employed, it would seem, as an optimization 
criterion. Distances between nodes 
in the dendrogram satisfy the (stronger than {\it triangular}) 
{\it ultrametric} 
inequality, $d_{ij} \leq \max{(d_{ik},d_{jk})}$ \cite[p. 245]{johnson} 
\cite[eq. (2.2)]{rammal}.

Geographic subdivisions (or groups of subdivisions) that enter into the 
bulk of the dendrogram at the weakest levels are those with the broadest ties. Typically, these have been found to be 
``cosmopolitan", hub-like areas, 
a prototypical example being 
the French capital, Paris \cite[sec. 4.1]{tree} \cite{france}. 
Similarly, 
in parallel analyses of other 
internal migration tables, the cosmopolitan/non-provincial natures
of London, Barcelona, Milan, West Berlin, Moscow, Manila, Bucharest, 
Montr{\'e}al, Z{\"u}rich, Santiago, Tunis and Istanbul were--among
others--highlighted in the respective dendrograms for their nations
 \cite[sec. 8.2]{tree} 
\cite[pp. 181-182]{qq} \cite[p. 55]{science}. In the intercounty analysis
for the US, the most cosmopolitan entities were: (1) the 
{\it centrally} located paired
Illinois counties of Cook (Chicago) and neighboring, suburban Du Page; 
(2) the nation's capital, Washington, D. C.; and (3) the paired south
Florida (retirement) counties of Dade (Miami) and Broward (Ft. Lauderdale) 
\cite{county,partial}. (In general, counties with large military 
installations, large college populations, or that were state capitals 
also interacted relatively broadly with other areas.)

It should be emphasized that 
although the indicated cosmopolitan areas may generally have 
relatively large populations, 
this can not, in and of itself, 
explain the wide national ties observed, since the 
double-standardization, in effect, renders all areas of equal overall size.
 
Additionally, 
geographically isolated areas--such as the Japanese islands of 
Kyushu and Shikoku--emerged as well-defined {\it clusters} (regions)
of their constituent subdivisions (``prefectures'' in the Japanese case) 
in the dendrograms (cf. \cite{e,multiterminal}), and similarly
 the Italian islands of Sicily and Sardinia 
\cite{gentileschi}, and the North and South Islands of New Zealand and 
Newfoundland (Canada) \cite[p. 1]{county}.
The eight counties of Connecticut, and other New England groupings, as  
further examples,  were
also very prominent in the highly disaggregated US analysis \cite{county}. 
Relatedly, in a study based solely upon 
the 1968 movement of {\it college students} among
the fifty states, the six New England states were strongly clustered 
\cite[Fig. 1]{college}.

Though quite successful, evidently, in revealing hub-like 
and clustering behavior in 
recorded flows, the indicated series of studies did not address the 
recently-emerging, theoretically-important 
issues of scale-free networks, power-law descriptions, network evolution 
and vulnerability, and 
small-world properties
that have been stressed by Barab{\'a}si 
\cite{linked} (and his colleagues and many others in the growing field). 
(For critiques of these matters, see \cite{doyle,alderson}.) 
In this regard, one might--using the indicated two-stage 
procedure--compare the hierarchical structure of geographic 
areas using internal migration tables at {\it different} levels of 
geographic aggregation
(counties, states, regions...)
To again use the example of France, based on a 
1962-68 $21 \times 21$ interregional
table, R{\'e}gion Parisienne was certainly the most hub-like 
\cite[sec. 4.1]{tree} \cite{france}, while using a finer
$89 \times 89$ 1954-62 interdepartmental table, the dyad 
composed of Seine 
(that is Paris and its immediate suburbs) together with the 
encircling
Seine-et-Oise (administratively eliminated in 1964) 
was most cosmopolitan \cite{winchester} 
\cite[sec. 6.1]{tree}.

It would be of interest to develop a theory--making use of 
the rich mathematical structure of doubly-stochastic 
matrices--by which the {\it statistical
significance} of apparent hubs and clusters 
in dendrograms could be evaluated \cite[pp. 7-8]{county} \cite{bock}. 
In the geographic context of internal migration tables, where nearby areas 
have a strong distance-adversion predilection for binding, it seems unlikely 
that most clustering
results generated could be considered to be--in any standard 
sense--``random'' in nature. 
On the other hand, other types of ``origin-destination'' 
tables, such as those for 
{\it occupational} mobility \cite{duncan}, journal citations 
\cite{science} \cite[pp. 125-153]{tree2}, interindustry (input-output) flows 
\cite{IO}, brand switches \cite[sec. 9.6]{tree} \cite{rao}, 
crime switches \cite[sec. 9.7]{tree} \cite[Table XII]{blumstein}, and (Morse code) 
confusions \cite[sec. 9.8]{tree} \cite{rothkopf}, among others, clearly lack such a geographic dimension. 
An efficient algorithm--considered as a nonlinear dynamical system--to generate {\it random} bistochastic matrices has
recently been presented \cite{CSBZ} (cf. \cite{griffiths,ZKSS}).

The creative, productive  network analyst M. E. J. Newman 
has written: ``Edge weights in networks have, with some exceptions
\ldots received relatively little attention in the {\it physics} 
[emphasis added]  literature for 
the excellent reason that in any field one is well advised to look at the
simple cases first (unweighted networks). On the other hand, there are 
many cases where edge weights are known for networks, and to ignore them
is to throw out a lot of data that, in theory at least, could help us to 
understand these systems better'' \cite{newman1}. Of course, the 
numerous applications of the two-stage 
procedure we have discussed 
above have, in fact, been to such weighted networks.

In \cite{newman1}, Newman applied the famous Ford-Fulkerson 
{\it max-flow/min-cut theorem} to
weighted networks (which he mapped onto unweighted {\it multigraphs}). Earlier, this theorem had been used to study 
Spanish \cite{multiterminal}, Philippine \cite{philippine}, and 
Brazilian, Mexican and Argentinian \cite{brazil} 
internal migration and 
US interindustry flows \cite[pp. 18-28]{tree2} \cite{IO2}--all the 
corresponding flows now being
left unadjusted, that is {\it not} standardized.
In this ``multiterminal'' 
approach, the maximum flow and the dual minimum edge cut-sets, 
between {\it all} ordered pairs of nodes are found. Those cuts 
(often few or even {\it null} in number) which partition the 
$N$ nodes nontrivially--that is, into two sets each of cardinality greater than
1--are noted. The set in each such pair with the fewer nodes is regarded
as a nodal cluster (region, in the geographic context). It has the 
interesting, defining property that fewer people migrate into (from) it, as 
a whole, than into (from) its node. In the Spanish context, the 
(nodal) province of 
Badajoz was found to have a particularly large out-migration sphere of
influence, and 
the (Basque) province of 
Vizcaya (site of Bilbao and Guernica), 
an extensive in-migration field \cite{multiterminal}.

The networks formed by the World Wide Web and the Internet have been the focus
of much recent interest \cite{linked}. Their structures are typically 
represented by 
$N \times N$ {\it adjacency} matrices, the entries of which are simply 0 or 1,
rather than natural numbers, as in internal migration and other
flow tables. One might investigate whether the two-stage
double-standardization and 
hierarchical clustering, and the (complementary) 
multiterminal max-flow/min-cut 
procedures we have sought to bring
to the attention of the active body of 
contemporary network theorists, could yield novel insights into these
 and other important modern structures.

In closing, it might be of interest to describe the 
immediate motivation for 
this particular note. I had done no further work applying the methods described above after 1985, being aware of, but not absorbed in recent
developments in network analysis. 
In May, 2008, Mathematical Reviews asked me to 
review the book of Tom Siegfried \cite{siegfried}, chapter 8 of which is
devoted to the on-going activities in network analysis. This further led 
me (thanks to D. E. Boyce) to the book of Barab{\' a}si \cite{linked}. 
I, then, e-mailed
Barab{\' a}si, pointing out the use of the earlier, widely-applied
clustering methods.
In reply, he wrote, in part: ``I guess you were another demo of everything being a question of timing-- after a quick look it does appear that many things you did have came back as questions -- with much more detailed data-- again in the network community today. No, I was not aware of your papers, 
unfortunately, and it is hard to know 
how to get them back into the flow of the system''. 
The present note might be seen as an effort in that direction, alerting
present-day investigators to these demonstratedly fruitful research 
methodologies.

\begin{acknowledgments}
I would like to express appreciation to the Kavli Institute for Theoretical Physics (KITP)
for technical support.
\end{acknowledgments}

\bibliography{Hub3}

\begin{thebibliography}{53}
\expandafter\ifx\csname natexlab\endcsname\relax\def\natexlab#1{#1}\fi
\expandafter\ifx\csname bibnamefont\endcsname\relax
  \def\bibnamefont#1{#1}\fi
\expandafter\ifx\csname bibfnamefont\endcsname\relax
  \def\bibfnamefont#1{#1}\fi
\expandafter\ifx\csname citenamefont\endcsname\relax
  \def\citenamefont#1{#1}\fi
\expandafter\ifx\csname url\endcsname\relax
  \def\url#1{\texttt{#1}}\fi
\expandafter\ifx\csname urlprefix\endcsname\relax\def\urlprefix{URL }\fi
\providecommand{\bibinfo}[2]{#2}
\providecommand{\eprint}[2][]{\url{#2}}

\bibitem[{\citenamefont{Barab{\'a}si}(2003)}]{linked}
\bibinfo{author}{\bibfnamefont{A.-L.} \bibnamefont{Barab{\'a}si}},
  \emph{\bibinfo{title}{Linked: How everything is connected to everything else
  and what it means for business, science, and everyday life}}
  (\bibinfo{publisher}{Plume}, \bibinfo{address}{New York},
  \bibinfo{year}{2003}).

\bibitem[{\citenamefont{Siegfried}(2006)}]{siegfried}
\bibinfo{author}{\bibfnamefont{T.}~\bibnamefont{Siegfried}},
  \emph{\bibinfo{title}{A beautiful math: John Nash, game theory, and the
  modern quest for a code of nature}} (\bibinfo{publisher}{Joseph Henry},
  \bibinfo{address}{Washington}, \bibinfo{year}{2006}).

\bibitem[{\citenamefont{Slater}(1984{\natexlab{a}})}]{tree}
\bibinfo{author}{\bibfnamefont{P.~B.} \bibnamefont{Slater}},
  \emph{\bibinfo{title}{Tree representations of internal migration flows and
  related topics}} (\bibinfo{publisher}{Community and Organization Res. Inst.},
  \bibinfo{address}{Santa Barbara}, \bibinfo{year}{1984}{\natexlab{a}}).

\bibitem[{\citenamefont{Dubes}(1985)}]{dubes}
\bibinfo{author}{\bibfnamefont{R.~C.} \bibnamefont{Dubes}},
  \bibinfo{journal}{J. Classif.} \textbf{\bibinfo{volume}{2}},
  \bibinfo{pages}{141} (\bibinfo{year}{1985}).

\bibitem[{\citenamefont{Slater}(1976{\natexlab{a}})}]{japan}
\bibinfo{author}{\bibfnamefont{P.~B.} \bibnamefont{Slater}},
  \bibinfo{journal}{Regional Stud.} \textbf{\bibinfo{volume}{10}},
  \bibinfo{pages}{123} (\bibinfo{year}{1976}{\natexlab{a}}).

\bibitem[{\citenamefont{Slater}(1976{\natexlab{b}})}]{france}
\bibinfo{author}{\bibfnamefont{P.~B.} \bibnamefont{Slater}},
  \bibinfo{journal}{IEEE Syst. Man. Cyb.} \textbf{\bibinfo{volume}{6}},
  \bibinfo{pages}{321} (\bibinfo{year}{1976}{\natexlab{b}}).

\bibitem[{\citenamefont{Slater and Winchester}(1978)}]{winchester}
\bibinfo{author}{\bibfnamefont{P.~B.} \bibnamefont{Slater}} \bibnamefont{and}
  \bibinfo{author}{\bibfnamefont{H.~L.~M.} \bibnamefont{Winchester}},
  \bibinfo{journal}{IEEE Syst. Man. Cyb.} \textbf{\bibinfo{volume}{8}},
  \bibinfo{pages}{635} (\bibinfo{year}{1978}).

\bibitem[{\citenamefont{Slater}(1983{\natexlab{a}})}]{science}
\bibinfo{author}{\bibfnamefont{P.~B.} \bibnamefont{Slater}},
  \bibinfo{journal}{Scientometrics} \textbf{\bibinfo{volume}{5}},
  \bibinfo{pages}{55} (\bibinfo{year}{1983}{\natexlab{a}}).

\bibitem[{\citenamefont{Slater}(1984{\natexlab{b}})}]{partial}
\bibinfo{author}{\bibfnamefont{P.~B.} \bibnamefont{Slater}},
  \bibinfo{journal}{Environ. Plann. A} \textbf{\bibinfo{volume}{16}},
  \bibinfo{pages}{545} (\bibinfo{year}{1984}{\natexlab{b}}).

\bibitem[{\citenamefont{Slater}(1977{\natexlab{a}})}]{IO}
\bibinfo{author}{\bibfnamefont{P.~B.} \bibnamefont{Slater}},
  \bibinfo{journal}{Empirical Econ.} \textbf{\bibinfo{volume}{2}},
  \bibinfo{pages}{1} (\bibinfo{year}{1977}{\natexlab{a}}).

\bibitem[{\citenamefont{Slater}(1976{\natexlab{c}})}]{college}
\bibinfo{author}{\bibfnamefont{P.~B.} \bibnamefont{Slater}},
  \bibinfo{journal}{Res. Higher Educ.} \textbf{\bibinfo{volume}{4}},
  \bibinfo{pages}{305} (\bibinfo{year}{1976}{\natexlab{c}}).

\bibitem[{\citenamefont{Gentileschi and Slater}(1980)}]{gentileschi}
\bibinfo{author}{\bibfnamefont{M.~L.} \bibnamefont{Gentileschi}}
  \bibnamefont{and} \bibinfo{author}{\bibfnamefont{P.~B.}
  \bibnamefont{Slater}}, \bibinfo{journal}{Riv. Geog. Ital.}
  \textbf{\bibinfo{volume}{87}}, \bibinfo{pages}{133} (\bibinfo{year}{1980}).

\bibitem[{\citenamefont{Slater}(1976{\natexlab{d}})}]{SEAS}
\bibinfo{author}{\bibfnamefont{P.~B.} \bibnamefont{Slater}},
  \bibinfo{journal}{Rev. Public Data Use} \textbf{\bibinfo{volume}{4}},
  \bibinfo{pages}{32} (\bibinfo{year}{1976}{\natexlab{d}}).

\bibitem[{\citenamefont{Slater}(1981)}]{qq}
\bibinfo{author}{\bibfnamefont{P.~B.} \bibnamefont{Slater}},
  \bibinfo{journal}{Quality and Quantity} \textbf{\bibinfo{volume}{15}},
  \bibinfo{pages}{179} (\bibinfo{year}{1981}).

\bibitem[{\citenamefont{Slater}(1983{\natexlab{b}})}]{county}
\bibinfo{author}{\bibfnamefont{P.~B.} \bibnamefont{Slater}},
  \emph{\bibinfo{title}{Migration regions of the United States: two
  county-level 1965-70 analyses}} (\bibinfo{publisher}{Community and
  Organization Res. Inst.}, \bibinfo{address}{Santa Barbara},
  \bibinfo{year}{1983}{\natexlab{b}}).

\bibitem[{\citenamefont{Slater}(1986)}]{tree2}
\bibinfo{author}{\bibfnamefont{P.~B.} \bibnamefont{Slater}},
  \emph{\bibinfo{title}{Large scale data analytic studies in the social
  sciences}} (\bibinfo{publisher}{Community and Organization Res. Inst.},
  \bibinfo{address}{Santa Barbara}, \bibinfo{year}{1986}).

\bibitem[{\citenamefont{Bacharach}(1970)}]{bacharach}
\bibinfo{author}{\bibfnamefont{M.~A.} \bibnamefont{Bacharach}},
  \emph{\bibinfo{title}{Biproportional matrices and input-output change}}
  (\bibinfo{publisher}{Cambridge Univ.}, \bibinfo{address}{Cambridge},
  \bibinfo{year}{1970}).

\bibitem[{\citenamefont{Mosteller}(1968)}]{mosteller}
\bibinfo{author}{\bibfnamefont{F.}~\bibnamefont{Mosteller}},
  \bibinfo{journal}{J. Amer. Statist. Assoc.} \textbf{\bibinfo{volume}{63}},
  \bibinfo{pages}{1} (\bibinfo{year}{1968}).

\bibitem[{\citenamefont{Louck}(1997)}]{louck}
\bibinfo{author}{\bibfnamefont{J.~D.} \bibnamefont{Louck}},
  \bibinfo{journal}{Found. Phys.} \textbf{\bibinfo{volume}{27}},
  \bibinfo{pages}{1085} (\bibinfo{year}{1997}).

\bibitem[{\citenamefont{Cappellini et~al.}()\citenamefont{Cappellini, Sommers,
  Bruzda, and \.Zyczkowski}}]{CSBZ}
\bibinfo{author}{\bibfnamefont{V.}~\bibnamefont{Cappellini}},
  \bibinfo{author}{\bibfnamefont{H.-J.} \bibnamefont{Sommers}},
  \bibinfo{author}{\bibfnamefont{W.}~\bibnamefont{Bruzda}}, \bibnamefont{and}
  \bibinfo{author}{\bibfnamefont{K.}~\bibnamefont{\.Zyczkowski}},
  \emph{\bibinfo{title}{Nonlinear dynamics in constructing random bistochastic
  matrices}}, \eprint{arXiv:0711.3345}.

\bibitem[{\citenamefont{Bengtsson}()}]{unistochastic}
\bibinfo{author}{\bibfnamefont{I.}~\bibnamefont{Bengtsson}},
  \emph{\bibinfo{title}{The importance of being unistochastic}},
  \eprint{quant-ph/0403088}.

\bibitem[{\citenamefont{Eriksson}(1980)}]{eriksson}
\bibinfo{author}{\bibfnamefont{J.}~\bibnamefont{Eriksson}},
  \bibinfo{journal}{Math. Program.} \textbf{\bibinfo{volume}{18}},
  \bibinfo{pages}{146} (\bibinfo{year}{1980}).

\bibitem[{\citenamefont{Parlett and Landis}(1982)}]{parlett}
\bibinfo{author}{\bibfnamefont{B.~N.} \bibnamefont{Parlett}} \bibnamefont{and}
  \bibinfo{author}{\bibfnamefont{T.~L.} \bibnamefont{Landis}},
  \bibinfo{journal}{Lin. Alg. Applics.} \textbf{\bibinfo{volume}{48}},
  \bibinfo{pages}{53} (\bibinfo{year}{1982}).

\bibitem[{\citenamefont{Sinkhorn and Knopp}(1967)}]{sinkhorn}
\bibinfo{author}{\bibfnamefont{R.}~\bibnamefont{Sinkhorn}} \bibnamefont{and}
  \bibinfo{author}{\bibfnamefont{P.}~\bibnamefont{Knopp}},
  \bibinfo{journal}{Pac. J. Math.} \textbf{\bibinfo{volume}{21}},
  \bibinfo{pages}{343} (\bibinfo{year}{1967}).

\bibitem[{\citenamefont{Mirsky}(1971)}]{mirsky}
\bibinfo{author}{\bibfnamefont{L.}~\bibnamefont{Mirsky}},
  \emph{\bibinfo{title}{Transversal Theory}} (\bibinfo{publisher}{Academic},
  \bibinfo{address}{New York}, \bibinfo{year}{1971}).

\bibitem[{\citenamefont{Simonoff}(1995)}]{simonoff}
\bibinfo{author}{\bibfnamefont{J.~S.} \bibnamefont{Simonoff}},
  \bibinfo{journal}{J. Statist. Plann. Infer.} \textbf{\bibinfo{volume}{47}},
  \bibinfo{pages}{41} (\bibinfo{year}{1995}).

\bibitem[{\citenamefont{Slater}(1980)}]{boundary}
\bibinfo{author}{\bibfnamefont{P.~B.} \bibnamefont{Slater}},
  \bibinfo{journal}{IEEE Syst. Man Cyber.} \textbf{\bibinfo{volume}{10}},
  \bibinfo{pages}{678} (\bibinfo{year}{1980}).

\bibitem[{\citenamefont{Linial et~al.}(2000)\citenamefont{Linial,
  Samorodnitsky, and Wigderson}}]{linial}
\bibinfo{author}{\bibfnamefont{N.}~\bibnamefont{Linial}},
  \bibinfo{author}{\bibfnamefont{A.}~\bibnamefont{Samorodnitsky}},
  \bibnamefont{and}
  \bibinfo{author}{\bibfnamefont{A.}~\bibnamefont{Wigderson}},
  \bibinfo{journal}{Combinatorica} \textbf{\bibinfo{volume}{20}},
  \bibinfo{pages}{545} (\bibinfo{year}{2000}).

\bibitem[{\citenamefont{Ozawa}(1983)}]{ozawa}
\bibinfo{author}{\bibfnamefont{K.}~\bibnamefont{Ozawa}},
  \bibinfo{journal}{Patt. Recog.} \textbf{\bibinfo{volume}{16}},
  \bibinfo{pages}{201} (\bibinfo{year}{1983}).

\bibitem[{\citenamefont{Leusmann}(1977)}]{leusmann}
\bibinfo{author}{\bibfnamefont{C.}~\bibnamefont{Leusmann}},
  \bibinfo{journal}{Comput. Applics.} \textbf{\bibinfo{volume}{769}},
  \bibinfo{pages}{769} (\bibinfo{year}{1977}).

\bibitem[{\citenamefont{Chilko}(1980)}]{chilko}
\bibinfo{author}{\bibfnamefont{D.}~\bibnamefont{Chilko}}, \bibinfo{journal}{SAS
  Supplemental Library User's Guide} pp. \bibinfo{pages}{65--70}
  (\bibinfo{year}{1980}).

\bibitem[{\citenamefont{Schwartz}(1982)}]{schwartz}
\bibinfo{author}{\bibfnamefont{J.}~\bibnamefont{Schwartz}},
  \bibinfo{journal}{Not. Amer. Math. Soc.} \textbf{\bibinfo{volume}{29}},
  \bibinfo{pages}{502} (\bibinfo{year}{1982}).

\bibitem[{\citenamefont{Tarjan}(1982)}]{tarjan}
\bibinfo{author}{\bibfnamefont{R.~E.} \bibnamefont{Tarjan}},
  \bibinfo{journal}{Info. Proc. Lett.} \textbf{\bibinfo{volume}{14}},
  \bibinfo{pages}{26} (\bibinfo{year}{1982}).

\bibitem[{\citenamefont{Tarjan}(1983)}]{tarjan2}
\bibinfo{author}{\bibfnamefont{R.~E.} \bibnamefont{Tarjan}},
  \bibinfo{journal}{Info. Proc. Lett.} \textbf{\bibinfo{volume}{17}},
  \bibinfo{pages}{37} (\bibinfo{year}{1983}).

\bibitem[{\citenamefont{Slater}(1987)}]{tarjanslater}
\bibinfo{author}{\bibfnamefont{P.~B.} \bibnamefont{Slater}},
  \bibinfo{journal}{Environ. Plann. A} \textbf{\bibinfo{volume}{19}},
  \bibinfo{pages}{117} (\bibinfo{year}{1987}).

\bibitem[{\citenamefont{Hartfiel and Spellman}(1972)}]{hartfiel}
\bibinfo{author}{\bibfnamefont{D.~F.} \bibnamefont{Hartfiel}} \bibnamefont{and}
  \bibinfo{author}{\bibfnamefont{J.~W.} \bibnamefont{Spellman}},
  \bibinfo{journal}{Proc. Amer. Math. Soc.} \textbf{\bibinfo{volume}{36}},
  \bibinfo{pages}{389} (\bibinfo{year}{1972}).

\bibitem[{\citenamefont{Johnson}(1967)}]{johnson}
\bibinfo{author}{\bibfnamefont{S.~C.} \bibnamefont{Johnson}},
  \bibinfo{journal}{Psychometrika} \textbf{\bibinfo{volume}{32}},
  \bibinfo{pages}{241} (\bibinfo{year}{1967}).

\bibitem[{\citenamefont{Rammal et~al.}(1986)\citenamefont{Rammal, Toulouse, and
  Virasoro}}]{rammal}
\bibinfo{author}{\bibfnamefont{R.}~\bibnamefont{Rammal}},
  \bibinfo{author}{\bibfnamefont{G.}~\bibnamefont{Toulouse}}, \bibnamefont{and}
  \bibinfo{author}{\bibfnamefont{M.~A.} \bibnamefont{Virasoro}},
  \bibinfo{journal}{Rev. Mod. Phys.} \textbf{\bibinfo{volume}{58}},
  \bibinfo{pages}{765} (\bibinfo{year}{1986}).

\bibitem[{\citenamefont{E et~al.}(2008)\citenamefont{E, Li, and
  Vanden-Eijnden}}]{e}
\bibinfo{author}{\bibfnamefont{W.}~\bibnamefont{E}},
  \bibinfo{author}{\bibfnamefont{T.}~\bibnamefont{Li}}, \bibnamefont{and}
  \bibinfo{author}{\bibfnamefont{E.}~\bibnamefont{Vanden-Eijnden}},
  \bibinfo{journal}{Proc. Natl. Acad. Sci} \textbf{\bibinfo{volume}{105}},
  \bibinfo{pages}{7907} (\bibinfo{year}{2008}).

\bibitem[{\citenamefont{Slater}(1976{\natexlab{e}})}]{multiterminal}
\bibinfo{author}{\bibfnamefont{P.~B.} \bibnamefont{Slater}},
  \bibinfo{journal}{Environ. Plann. A} \textbf{\bibinfo{volume}{8}},
  \bibinfo{pages}{875} (\bibinfo{year}{1976}{\natexlab{e}}).

\bibitem[{\citenamefont{Doyle et~al.}(2005)\citenamefont{Doyle, Alderson, Li,
  Low, Roughan, Shalunov, Tanaka, and Willinger}}]{doyle}
\bibinfo{author}{\bibfnamefont{J.~C.} \bibnamefont{Doyle}},
  \bibinfo{author}{\bibfnamefont{D.~L.} \bibnamefont{Alderson}},
  \bibinfo{author}{\bibfnamefont{L.}~\bibnamefont{Li}},
  \bibinfo{author}{\bibfnamefont{S.}~\bibnamefont{Low}},
  \bibinfo{author}{\bibfnamefont{M.}~\bibnamefont{Roughan}},
  \bibinfo{author}{\bibfnamefont{S.}~\bibnamefont{Shalunov}},
  \bibinfo{author}{\bibfnamefont{R.}~\bibnamefont{Tanaka}}, \bibnamefont{and}
  \bibinfo{author}{\bibfnamefont{W.}~\bibnamefont{Willinger}},
  \bibinfo{journal}{Proc. Natl. Acad. Sci.} \textbf{\bibinfo{volume}{102}},
  \bibinfo{pages}{14497} (\bibinfo{year}{2005}).

\bibitem[{\citenamefont{Alderson et~al.}(2005)\citenamefont{Alderson, Doyle,
  Li, and Willinger}}]{alderson}
\bibinfo{author}{\bibfnamefont{D.}~\bibnamefont{Alderson}},
  \bibinfo{author}{\bibfnamefont{J.~C.} \bibnamefont{Doyle}},
  \bibinfo{author}{\bibfnamefont{L.}~\bibnamefont{Li}}, \bibnamefont{and}
  \bibinfo{author}{\bibfnamefont{W.}~\bibnamefont{Willinger}},
  \bibinfo{journal}{Internet Math.} \textbf{\bibinfo{volume}{2}},
  \bibinfo{pages}{421} (\bibinfo{year}{2005}).

\bibitem[{\citenamefont{Bock}(1996)}]{bock}
\bibinfo{author}{\bibfnamefont{H.~H.} \bibnamefont{Bock}},
  \bibinfo{journal}{Comput. Stat. Data Anal.} \textbf{\bibinfo{volume}{23}},
  \bibinfo{pages}{5} (\bibinfo{year}{1996}).

\bibitem[{\citenamefont{Duncan}(1979)}]{duncan}
\bibinfo{author}{\bibfnamefont{O.~D.} \bibnamefont{Duncan}},
  \bibinfo{journal}{Amer. J. Sociol.} \textbf{\bibinfo{volume}{84}},
  \bibinfo{pages}{793} (\bibinfo{year}{1979}).

\bibitem[{\citenamefont{Rao and Sabavala}(1981)}]{rao}
\bibinfo{author}{\bibfnamefont{V.~R.} \bibnamefont{Rao}} \bibnamefont{and}
  \bibinfo{author}{\bibfnamefont{D.~J.} \bibnamefont{Sabavala}},
  \bibinfo{journal}{J. Consumer Res.} \textbf{\bibinfo{volume}{8}},
  \bibinfo{pages}{85} (\bibinfo{year}{1981}).

\bibitem[{\citenamefont{Blumstein and Larson}(1969)}]{blumstein}
\bibinfo{author}{\bibfnamefont{A.}~\bibnamefont{Blumstein}} \bibnamefont{and}
  \bibinfo{author}{\bibfnamefont{R.}~\bibnamefont{Larson}},
  \bibinfo{journal}{Operat. Res.} \textbf{\bibinfo{volume}{17}},
  \bibinfo{pages}{199} (\bibinfo{year}{1969}).

\bibitem[{\citenamefont{Rothkopf}(1957)}]{rothkopf}
\bibinfo{author}{\bibfnamefont{E.~Z.} \bibnamefont{Rothkopf}},
  \bibinfo{journal}{J. Experiment. Psych.} \textbf{\bibinfo{volume}{53}},
  \bibinfo{pages}{94} (\bibinfo{year}{1957}).

\bibitem[{\citenamefont{Griffiths}(1974)}]{griffiths}
\bibinfo{author}{\bibfnamefont{R.~C.} \bibnamefont{Griffiths}},
  \bibinfo{journal}{Canad. J. Math.} \textbf{\bibinfo{volume}{26}},
  \bibinfo{pages}{600} (\bibinfo{year}{1974}).

\bibitem[{\citenamefont{{\.Z}yczkowski
  et~al.}(2003)\citenamefont{{\.Z}yczkowski, Ku{\'s}, S{\l}omczy{\'n}ski, and
  Sommers}}]{ZKSS}
\bibinfo{author}{\bibfnamefont{K.}~\bibnamefont{{\.Z}yczkowski}},
  \bibinfo{author}{\bibfnamefont{M.}~\bibnamefont{Ku{\'s}}},
  \bibinfo{author}{\bibfnamefont{W.}~\bibnamefont{S{\l}omczy{\'n}ski}},
  \bibnamefont{and} \bibinfo{author}{\bibfnamefont{H.-J.}
  \bibnamefont{Sommers}}, \bibinfo{journal}{J. Phys. A}
  \textbf{\bibinfo{volume}{36}}, \bibinfo{pages}{3425} (\bibinfo{year}{2003}).

\bibitem[{\citenamefont{Newman}(2004)}]{newman1}
\bibinfo{author}{\bibfnamefont{M.~E.~J.} \bibnamefont{Newman}},
  \bibinfo{journal}{Phys. Rev. E} \textbf{\bibinfo{volume}{70}},
  \bibinfo{pages}{056131} (\bibinfo{year}{2004}).

\bibitem[{\citenamefont{Slater}(1976{\natexlab{f}})}]{philippine}
\bibinfo{author}{\bibfnamefont{P.~B.} \bibnamefont{Slater}},
  \bibinfo{journal}{Philippine Geog. J.} \textbf{\bibinfo{volume}{20}},
  \bibinfo{pages}{79} (\bibinfo{year}{1976}{\natexlab{f}}).

\bibitem[{\citenamefont{Slater}(1977{\natexlab{b}})}]{brazil}
\bibinfo{author}{\bibfnamefont{P.~B.} \bibnamefont{Slater}},
  \bibinfo{journal}{Estad{\'i}stica} \textbf{\bibinfo{volume}{36}},
  \bibinfo{pages}{180} (\bibinfo{year}{1977}{\natexlab{b}}).

\bibitem[{\citenamefont{Slater}(1977{\natexlab{c}})}]{IO2}
\bibinfo{author}{\bibfnamefont{P.~B.} \bibnamefont{Slater}},
  \bibinfo{journal}{Empirical Econ.} \textbf{\bibinfo{volume}{3}},
  \bibinfo{pages}{49} (\bibinfo{year}{1977}{\natexlab{c}}).

\end{thebibliography}

\end{document}